\documentclass[10pt,a4paper]{article}

\usepackage[utf8]{inputenc}
\usepackage[english]{babel}
\usepackage{amssymb,amsmath}
\usepackage{euscript}

\unitlength=1.00mm\linethickness{0.4pt}
\begin{document}
\large

\newpage
\begin{center}
\LARGE{\bf An Allgravity as a Grand Unification of Forces}
\end{center}
\vspace{0.1mm}
\begin{center}
{\bf Rasulkhozha S. Sharafiddinov}
\end{center}
\vspace{0.1mm}
\begin{center}
{\bf Institute of Nuclear Physics, Uzbekistan Academy of Sciences,
\\Ulugbek, Tashkent 100214, Uzbekistan}
\end{center}
\vspace{0.1mm}

\begin{center}
{\bf Abstract}
\end{center}

Each type of Coulomb (Newton) charge corresponds to a kind of Coulomb (Newton) mass. Such a 
mass-charge duality principle explains the availability of the united rest mass and charge in a neutrino equal to all its mass and charge consisting of the electric, weak, and strong components 
and the range of other innate components. A neutrino itself, similarly to all other quantum matter with Coulomb (Newton) mass and charge, testifies hereby in favor of a kind of mononeutrino with magnetocoulomb (magnetonewton) rest mass and charge equal to all its mass and charge including the magnetoelectric, magnetoweak, and magnetostrong parts and the range of other innate parts. We discuss a theory in which symmetry between electricity and magnetism comes forward at the level of a grand unification mathematical logic as the defined symmetry between gravity and magnetogravity within the same allgravity responsible for all that in a curved space-time. This allgravity relates a graviton 
and a monograviton as a consequence of force unification, forming a single allgraviton. Thereby, it establishes a set of forces and the role of mass and charge in their formation and thus directly reveals the most diverse properties of a curved space that have remained hitherto latent. 

\vspace{0.3cm}

\vspace{0.6cm}
\noindent
{\bf 1. Introduction}
\vspace{0.3cm}

Nature has been created such that each type of charge corresponds to a kind of inertial mass. This correspondence principle expresses the mass-charge duality [1], which confirms that we cannot exclude 
the coexistence of the naturally united rest mass $m_{\nu}^{U}$ and charge $e_{\nu}^{U},$ for example, in a neutrino $\nu,$ equal to all its mass and charge
\begin{equation}
m_{\nu}=m_{\nu}^{U}=m_{\nu}^{E}+m_{\nu}^{W}+m_{\nu}^{S}+...,
\label{1}
\end{equation}
\begin{equation}
e_{\nu}=e_{\nu}^{U}=e_{\nu}^{E}+e_{\nu}^{W}+e_{\nu}^{S}+...,
\label{2}
\end{equation}
each of which includes the electric $(E),$ weak $(W),$ and strong $(S)$ contributions and 
the range of other innate contributions.

All neutrinos, similarly to all other quantum matter, hereby have a gravitational mass that 
corresponds to its gravitational charge. A distinction of their own values from the inertial 
sizes of the corresponding pair of the naturally united $(U)$ rest mass and charge implies the availability between the objects of the interaction fifth force [2], which comes forward 
in a system as a unified whole.

Another possibility, in principle, is that regardless of the maximal quantity of forces and existing types of actions, all inertial masses and charges are strictly gravitational ones. At the same time, the structure of gravity remains an open question. 

It is usually assumed that the gravitational force cannot appear at the microworld level 
because of its weakness. There exists simultaneously a real possibility of observing not only  massive neutrinos [3] or photons [4-6] but also the gravitomagnetic (magnetogravitational) force 
of the Earth [7-9], which could not be conducted without the presence of that latently acts on all matter without exception. Therefore, a subtle measurement of the rest mass of the neutrino [10,11] 
or the photon [12] must be considered as one of the available laboratory data sets confirming the existence of microgravity. No one is hereby in force to define the fundamental interaction structure of elementary objects regardless of gravity.

Here we investigate whether there exists any defined symmetry between gravity and magnetogravity 
and, if so, what the expected symmetrical structure indicates about the unification of forces of 
a different nature. This, in turn, requires elucidating the idea of each mass-charge pair of the existing types of masses and charges, which may serve as the source of a kind of naturally united gauge field. Insofar as a logically consistent mathematical formulation of the suggested field 
theory of grand unification is concerned, a single interaction is responsible for all that is connected with its wave equations in a curved space-time.

\vspace{0.6cm}
\noindent
{\bf 2. The unified mass-charge structure of fundamental forces}
\vspace{0.3cm}

According to the mass-charge duality [1], the Newton force of gravitation $F_{N_{\nu\nu}}$ and  Coulomb force $F_{C_{\nu\nu}}$ between two neutrinos [13] must be defined from the point of view of any of the existing types of actions. In other words, each type of force includes a kind of Newton contribution and a kind of Coulomb contribution [2].

To conform with these implications, we not only conclude that
\begin{equation}
m_{\nu}=m_{\nu}^{U}=m_{\nu}^{G}=m_{\nu}^{GU}=
m_{\nu}^{E}+m_{\nu}^{W}+m_{\nu}^{S}+...,
\label{3}
\end{equation}
\begin{equation}
e_{\nu}=e_{\nu}^{U}=e_{\nu}^{G}=e_{\nu}^{GU}=
e_{\nu}^{E}+e_{\nu}^{W}+e_{\nu}^{S}+...
\label{4}
\end{equation}
but also must recognize that gravitation comes forward in nature as a unification. The gravitational $(G)$ mass and charge of a neutrino are hereby the naturally united $(GU)$ mass and charge equal to all its mass and charge, each of which consists of electric, weak, and strong parts and the range of other innate parts. 

For our purposes, it is desirable to first present the Newton and Coulomb forces explained 
by the corresponding masses and charges of each pair of interacting objects in a general 
$(K=E,$ $W,$ $S,...)$ form in the following manner:
\begin{equation}
F^{K}_{N_{\nu\nu}}=
G_{N}\left(\frac{m_{\nu}^{K}}{r}\right)^{2}, \, \, \, \,
F^{K}_{C_{\nu\nu}}=\frac{1}{4\pi\epsilon_{0}}
\left(\frac{e_{\nu}^{K}}{r}\right)^{2},
\label{5}
\end{equation}
\begin{equation}
F^{ij}_{N_{\nu\nu}}=
G_{N}\frac{m_{\nu}^{i}m_{\nu}^{j}}{r^{2}}, \, \, \, \,
F^{ij}_{C_{\nu\nu}}=\frac{1}{4\pi\epsilon_{0}}
\frac{e_{\nu}^{i}e_{\nu}^{j}}{r^{2}},
\label{6}
\end{equation}
where $i,$ $j=K$ $(i\neq j),$ $r$ denotes the distance between the particles, and $G_{N}$ is a 
gravitational constant.

Exactly the same one can define the structure of the studied forces for each pair of naturally 
united $(GU)$ masses and charges:
\begin{equation}
F^{GU}_{N_{\nu\nu}}=
G_{N}\left(\frac{m_{\nu}^{GU}}{r}\right)^{2}, \, \, \, \,
F^{GU}_{C_{\nu\nu}}=\frac{1}{4\pi\epsilon_{0}}
\left(\frac{e_{\nu}^{GU}}{r}\right)^{2}.
\label{7}
\end{equation}

Inserting (\ref{3}) and (\ref{4}) in (\ref{7}), uniting the findings with (\ref{5}) and (\ref{6}) 
and having in mind the equalities
\begin{equation}
F^{K}_{\nu\nu}=F^{K}_{N_{\nu\nu}}+F^{K}_{C_{\nu\nu}},
\label{8}
\end{equation}
\begin{equation}
F^{ij}_{\nu\nu}=F^{ij}_{N_{\nu\nu}}+F^{ij}_{C_{\nu\nu}},
\label{9}
\end{equation}
we are led to the implication that
\begin{equation}
F^{GU}_{N_{\nu\nu}}=
F^{E}_{N_{\nu\nu}}+F^{W}_{N_{\nu\nu}}+F^{S}_{N_{\nu\nu}}+...,
\label{10}
\end{equation}
\begin{equation}
F^{GU}_{C_{\nu\nu}}=
F^{E}_{C_{\nu\nu}}+F^{W}_{C_{\nu\nu}}+F^{S}_{C_{\nu\nu}}+....
\label{11}
\end{equation}

These solutions and 
\begin{equation}
F^{GU}_{\nu\nu}=F^{GU}_{N_{\nu\nu}}+F^{GU}_{C_{\nu\nu}}
\label{12}
\end{equation}
allow one to conclude that
\begin{equation}
F^{GU}_{\nu\nu}=
F^{E}_{\nu\nu}+F^{W}_{\nu\nu}+ F^{S}_{\nu\nu}+...,
\label{13}
\end{equation}
which leads to the appearance of naturally united gauge fields of matter. Such a regularity takes 
place because of the unified mass-charge structure of all types of forces of nonmagnetic behavior. They hereby constitute the interaction united force, which comes forward in nature as gravity.

Insofar as the magnetic forces are concerned, we will start here from the requirement [14] 
that the possibility of the existence of Dirac fermions simultaneously having electric $(E)$ and
magnetic $(H)$ charges is not excluded. To include the magnetic forces in the discussion, one must rewrite (\ref{1})-(\ref{4}) to account for the availability of magnetic mass  $m_{\nu}^{H}$ and charge $e_{\nu}^{H}$ in a neutrino. This gives the right to replace (\ref{13}) with
\begin{equation}
F^{GU}_{\nu\nu}=
F^{EH}_{\nu\nu}+F^{W}_{\nu\nu}+ F^{S}_{\nu\nu}+...
\label{14}
\end{equation}
in which appears a contribution of the united electromagnetic $(EH)$ force
\begin{equation}
F^{EH}_{\nu\nu}=F^{E}_{\nu\nu}+F^{H}_{\nu\nu}
\label{15}
\end{equation}
including a Newton-Coulomb pair of magnetic forces 
\begin{equation}
F^{H}_{\nu\nu}=F^{H}_{N_{\nu\nu}}+F^{H}_{C_{\nu\nu}}.
\label{16}
\end{equation}

So, in principle, each of the above forces $F^{K}_{\nu\nu}$ that contains a kind of Newton component
and a kind of Coulomb component has the unified mass-charge structure. They can therefore be called the gravitoelectric $(G_{E}),$ magnetogravitational $(H_{G}),$ electromagnetogravitational $(EH_{G}),$ gravitoweak $(G_{W})$ and gravitostrong $(G_{S})$ forces. It is also relevant that the names $(K=E,$ $W,$ $S,...)$ of these types of forces follow from historical traditions, not excepting that none of them had the characteristic of any freedom. 

\vspace{0.6cm}
\noindent
{\bf 3. Graviton field and its gauge nature}
\vspace{0.3cm}

There exists one more possibility [15], in principle, that any electrically charged particle 
testifies in favor of the existence of a kind of magnetically charged monoparticle possessing 
the magnetoelectric mass and charge. Such a sight on the nature of the fundamental symmetry 
between electricity and magnetism explains the fact that the presence of the photon $\gamma_{E}$ 
with electric $(E)$ mass [4-6] and charge [16] implies the existence of a kind of monophoton $\gamma_{H_{E}}$ with magnetoelectric $(H_{E})$ mass and charge. From this point of view, an electromagnetic field $\{\vec{E}, \vec{H_{E}}\}$ must arise as the field of the unified system
of the photon and monophoton $\{\gamma_{E}, \gamma_{H_{E}}\},$ where two Newton-Coulomb pairs
of forces of electric and magnetoelectric nature are united. We can therefore call the latter 
an electromagnetic $(EH_{E})$ boson:
\begin{equation}
\gamma_{EH_{E}}=\{\gamma_{E}, \, \, \, \, \gamma_{H_{E}}\}.
\label{17}
\end{equation}
This structural photon and the mediating particles $(W)$ of the weak interactions
constitute an electroweak $(EW)$ boson:
\begin{equation}
\gamma_{EW}=\{\gamma_{EH_{E}}, \, \, \, \, W\}
\label{18}
\end{equation}
if this is not forbidden by grand unification laws.

These facts and all that neutrino mass and charge imply about the structure of gravity imply that a gravitational field is strictly a united field of the unified system of the most diverse combinations of photons $(\gamma_{E}),$ monophotons $(\gamma_{H_{E}}),$ weak bosons $(W),$ and strong gluons 
$(g),$ where the four pairs of fundamental forces are united. In it appears the complicated 
structure of the graviton 
\begin{equation}
\gamma_{GU}=\{\gamma_{EH_{E}}, \, \, \, \, W, \, \, \, \, g\},
\label{19}
\end{equation}
which can be called the united gauge boson.

Returning to the neutrino, we remark that in the presence of a mass, it must possess [17] each of 
the possible types of charges and moments. This connection between the mass of a particle and its nature can explain the existence of the gravitational field not only of charge [13] or magnetic moment [18] but also of any type of current. 

It is not excluded, however, that from the point of view of mass-charge duality [1], each of the existing types of charges comes forward in a neutrino as the source of a kind of dipole moment [19] connected with the corresponding component of its mass. Therefore, if any of the photon and other  intragraviton boson currents consists of the Coulomb and Newton parts that correspond to the coexistence of mass and charge structures [20,21] of the gauge invariance, this will indicate 
the gauge behavior of the graviton field.

\vspace{0.6cm}
\noindent
{\bf 4. Nature of the curvature of space-time}
\vspace{0.3cm}

At first sight, a graviton as the united gauge boson [22] may be described by a vector (axial-vector) field, but this is not quite so. The point is that the vectors (axial vectors) constitute the plane surface. The same corresponding individual force of a Newton or Coulomb nature does not influence 
a given procedure, and consequently, the field of such an action remains linear. Unlike this, the electric, weak, and strong parts and other innate parts of gravitational force arise because of the harmony of a kind of Newton-Coulomb pair of forces. However, at the interratio of each Newton-Coulomb pair of intragraviton forces, their individual vectors (axial vectors) become naturally curved vectors (axial vectors), which constitute the curved surface.

It is already clear from the foregoing that a curvature of space-time reflects the availability 
in it of the harmony of forces of a different nature. Therefore, an individual field of each of the gauge bosons must be considered as the naturally curved field of the unified system of the two fields of the Coulomb and Newton behavior in which appears a part of the mass-charge structure [20,21] 
of gauge invariance. This becomes more interesting if we include in the discussion an interratio of intranuclear forces. Their harmony responds to the periodic revolution around the nucleus of any of the corresponding types of fermions of leptonic families [23]. At such a situation, the field of action of each of the structural components of the gravitational force between the nucleus and its satellite must be naturally curved. As a consequence, each lepton with a linear velocity in an atom undergoes orbital rather than straight-line motion. In them, some latent connections consequently appear in the mass-charge structure [20,21] dependence of the united gauge invariance of an intraatomic unified force. They hereby define the behavior of the structural objects of an 
atom at a true quantum level.

\vspace{0.6cm}
\noindent
{\bf 5. Mirror antisymmetry and current structure of gravity}
\vspace{0.3cm}

One of the predictions of the discussed theory of unification is the existence in nature of antigravity. From its point of view, an antigravity [24-26] reflects at a latent quantum level 
the characteristic features of electric, weak, and strong matter and the range of other innate types of matter because they constitute the naturally united gravity. In other words, the question about the invariance of the gravitational interaction concerning $C,$ $P$ and $T,$ and their combinations $CP$ and $CPT$ is intimately connected with the question about the behavior of its structural parts.

In principle, this circumstance requires one to consider why neutrino mass leads to the origin of an interconversion $\nu_{L}\leftrightarrow \nu_{R} ({\bar \nu_{R}}\leftrightarrow {\bar \nu_{L}}),$ for example, in polarized neutrino (antineutrino) scattering [17,27] by nuclei. However, in the same 
form as suggested, the standard electroweak model [28-30] is not in a state to describe the nature 
of these transitions between the left $(L)$ and right $(R),$ which violate the $P$-parity of neutrinos without a change in their lepton flavors. 

According to our description, the $P$-invariance of a particle is basically spontaneously violated 
at the expense of its rest mass [31]. Therefore, a mirror symmetry can exist only when gravity 
is absent.

Thus, if nature itself is not in force to create a true picture of the microworld fundamental interactions regardless of gravity, the suggested field theory of unification is strictly the 
left-right antisymmetrical theory, which describes the left- and right-handed particles taking 
into account that the same lepton cannot be simultaneously a left- and a right-handed fermion. 
Therefore, it predicts an equality of the number of components in all types of fermions [32].

Concerning the form of the united interaction, we gave the above reasoning that each of the electric, weak, and strong gauge bosons and the range of other innate types of gauge bosons comes forward in a graviton as the source of a kind of curved field, which constitutes one component of a gravitational field. Thereby, it plays an important role in establishing the fundamental symmetry between the unified graviton field and its structural parts. On this basis, the current structure of the naturally united gravitational interaction appears.

At first sight, this conclusion is not standard. However, the number of components in all types of forces coincide [2], as follows from considerations of symmetry. At the same time, nature itself has simultaneously the electric, weak, and strong Coulomb (Newton) interactions and the range of other innate types of Coulomb (Newton) interactions. This becomes possible because their unified behavior allows them to constitute naturally united gravity.

\vspace{0.6cm}
\noindent
{\bf 6. Coulomb structure of gauge invariance}
\vspace{0.3cm}

One of the most highlighted features of the first-initial Newton-Coulomb unification is its 
notion of Newton-Coulomb pairs of gauge bosons, each of which comes forward in nature as a single mediating particle of one of the structural components $(K=E,$ $W,$ $S,...)$ of the naturally united gravitational interaction. Its existence, as will be seen, must also be accepted as a consequence of the invariance of the corresponding Dirac Lagrangian concerning a kind of pair of local 
gauge transformations. 

To reveal this feature, one must refer to the free Dirac Lagrangian because it, with the use of the matrices $\gamma^{\mu}$ and neutrinos of mass $m_{\nu}^{K}$ and the four-component wave function $\psi_{\nu}(t, {\bf x}),$ may have the following structure: 
\begin{equation}
L_{free}^{D}=\overline{\psi}_{\nu}(i\gamma^{\mu}\partial_{\mu}^{K}-m_{\nu}^{K})\psi_{\nu}.
\label{20}  
\end{equation} 

Of course, this Lagrangian with quantum operator $\partial_{\mu}^{K},$ which expresses the idea 
of the charge structure [21] of gauge invariance, is invariant regarding the gauge transformation 
of a Coulomb $(C)$ nature
\begin{equation}
\psi'_{\nu}(x)=U_{K}^{C}\psi_{\nu}(x), \, \, \, \, U_{K}^{C}=e^{i\alpha_{K}(x)}
\label{21}  
\end{equation} 
if $\alpha_{K}$ does not depend on the space-time coordinates and becomes a global phase, in which 
case it is a global gauge transformation.

In the presence of a Coulomb transformation with the local phase $\alpha_{K}(x),$ the expected structure of the Lagrangian (\ref{20}) acquires the appearance of the structural term and requires 
the restoration of its broken internal symmetry from the point of view of the dynamical origin of the charge of fermions. Consequently, we must first introduce a Coulomb field $A_{\mu}^{K}(x)$ such that
it admits the existence of (\ref{21}) as a gauge transformation
\begin{equation}
A_{\mu}^{K'}=A_{\mu}^{K}+\frac{i}{e_{\nu}^{K}}\partial_{\mu}^{K}\alpha_{K}
\label{22}  
\end{equation} 
with a Coulomb constant $e_{\nu}^{K}$ at the level of a C-invariant charge of any $(K)$ nature. Therefore, the quantum operator $\partial_{\mu}^{K}$ with Coulomb behavior may be written as
\begin{equation}
\partial_{\mu}^{K}=\partial_{\mu}^{K}-e_{\nu}^{K}A_{\mu}^{K}.
\label{23}  
\end{equation} 

Thus, we have the possibility, in principle, on the basis of (\ref{23}), of establishing an invariance of the Dirac Lagrangian 
$$L^{D}=L_{free}^{D}+L_{int}^{D}=$$
\begin{equation}
=\overline{\psi}_{\nu}(i\gamma^{\mu}\partial_{\mu}^{K}-m_{\nu}^{K})\psi_{\nu}-
ie_{\nu}^{K}\overline{\psi}_{\nu}\gamma^{\mu}\psi_{\nu}A_{\mu}^{K}
\label{24}  
\end{equation} 
concerning the action of the chosen pair of local gauge transformations (\ref{21}) and (\ref{22}) 
due to an interaction of all types of neutrinos with the emission field of the photon. 

However, as stated in (\ref{24}), the field $A_{\mu}^{K}$ equalized with the Coulomb field is not of the same fields, the source of which may serve an object with an electric charge. In other words, the appearance of any of the gauge fields $A_{\mu}^{K}$ in a Lagrangian (\ref{24}) at $K=E,$ $W,$ $S,...$ indicates the availability in each of the existing types of interactions of a kind of mediating boson
of a Coulomb nature. It comes forward as one of the gauge bosons such that each of them gives a kind of gravitational charge to all objects that interact with its matter field.

This principle is not changed even at $K\rightarrow H_{K},$ $C\rightarrow H_{C},$ 
$\nu\rightarrow \nu_{H},$ and consequently, each of the magnetoelectric $(H_{E}),$ 
magnetoweak $(H_{W}),$ and magnetostrong $(H_{S})$ parts and the range of other innate parts of the magnetogravitational $(H_{G})$ charge of an elementary monoparticle is a consequence of a kind of interaction between this monoparticle and one of the corresponding types of gauge monobosons $(A_{\mu}^{H_{K}})$ having a magnetocoulomb $(H_{C})$ nature. 

The charges of a mononeutrino $\nu_{H}$ hereby constitute the united $(U_{H})$ magnetogravitational $(H_{G})$ charge equal to its grand magnetounited $(GU_{H})$ charge 
\begin{equation}
e_{\nu_{H}}=e_{\nu_{H}}^{U_{H}}=e_{\nu_{H}}^{H_{G}}=e_{\nu_{H}}^{GU_{H}}=
e_{\nu_{H}}^{H_{E}}+e_{\nu_{H}}^{H_{W}}+e_{\nu_{H}}^{H_{S}}+...
\label{25}
\end{equation}
defining the magnetocoulomb forces between two mononeutrinos as
\begin{equation}
F^{H_{K}}_{H_{C}^{\nu_{H}\nu_{H}}}=\frac{1}{4\pi\epsilon_{0}}
\left(\frac{e_{\nu_{H}}^{H_{K}}}{r_{H}}\right)^{2}, \, \, \, \,
F^{H_{i}H_{j}}_{H_{C}^{\nu_{H}\nu_{H}}}=\frac{1}{4\pi\epsilon_{0}}
\frac{e_{\nu_{H}}^{H_{i}}e_{\nu_{H}}^{H_{j}}}{r_{H}^{2}},
\label{26}
\end{equation}
where $H_{i},$ $H_{j}=H_{K}$ $(H_{i}\neq H_{j}),$ and $r_{H}$ denotes the distance 
among the monoparticles.

\vspace{0.6cm}
\noindent
{\bf 7. The unified Newton-Coulomb structure of gauge invariance}
\vspace{0.3cm}

The preceding reasoning says that nature itself can characterize each particle by its space-time coordinates $(t, {\bf x})$ if and only if its lifetime $\tau$ in the space where it exists does 
not exclude this. Therefore, without loss of unity of the principles of quantum mechanics, $m_{\nu}^{K},$ similarly to each of 
\begin{equation}
E_{\nu}^{K}=i\frac{\partial}{\partial t}, \, \, \, \,  
{\bf p}_{\nu}^{K}=-i\frac{\partial}{\partial {\bf x}}, \, \, \, \,  
\partial_{\mu}^{K}=
\frac{\partial}{\partial{\it x}^{\mu}}=\left(\frac{\partial}{\partial t}, 
-\frac{\partial}{\partial {\bf x}}\right), 
\label{27}
\end{equation}
must be a quantum operator [31] 
\begin{equation}
m_{\nu}^{K}=-i\frac{\partial}{\partial \tau} 
\label{28}
\end{equation}
such that
\begin{equation}
L_{free}^{D}=i\overline{\psi}_{\nu}(\gamma^{\mu}\partial_{\mu}^{K}+\partial_{\tau}^{K})\psi_{\nu}
\label{29}  
\end{equation} 
includes a kind of Coulomb $\partial_{\mu}^{K}$ component and a kind of Newton $\partial_{\tau}^{K}$ 
component. 

Therefore, we must recognize that the invariance of the free Dirac Lagrangian under the action of 
one more another type of local gauge transformation admits the appearance in it of the corresponding Newton field of one of the structural particles of one of the abovementioned Newton-Coulomb pairs 
of gauge bosons. 

Of course, this second type of transformation can appear in the mass structure [21] dependence 
of gauge invariance and has a Newton $(N)$ nature
\begin{equation}
\psi'_{\nu}=U_{K}^{N}\psi_{\nu}, \, \, \, \, U_{K}^{N}=e^{i\alpha_{K}(\tau)}.
\label{30}  
\end{equation} 

An internal feature of both types of transformations (\ref{21}) and (\ref{30}) is their 
coexistence. As a consequence, one of them says in favor of the second. They hereby constitute 
the Newton-Coulomb pairs of gauge transformations. Therefore, the operators $\partial_{\mu}^{K}$ 
and $\partial_{\tau}^{K}$ are such that
\begin{equation}
\partial_{\mu}^{K}\psi_{\nu}=\partial_{\mu}^{K}\psi_{\nu}(x), \, \, \, \, 
\partial_{\tau}^{K}\psi_{\nu}=\partial_{\tau}^{K}\psi_{\nu}(\tau),
\label{31}
\end{equation}
\begin{equation}
\partial_{\mu}^{K}\alpha_{K}=\partial_{\mu}^{K}\alpha_{K}(x), \, \, \, \, 
\partial_{\tau}^{K}\alpha_{K}=\partial_{\tau}^{K}\alpha_{K}(\tau).
\label{32}
\end{equation}

In the presence of a Newton transformation with the local phase $\alpha_{K}(\tau),$ the expected structure of the Lagrangian (\ref{29}) encounters the appearance of one more structural term and requires the restoration of its broken internal symmetry from the point of view of the dynamical origin of the mass of fermions. Thus, we must now introduce a Newton field $A_{\tau}^{K}(\tau)$
such that it gives the availability of (\ref{30}) as a gauge transformation
\begin{equation}
A_{\tau}^{K'}=A_{\tau}^{K}+\frac{i}{m_{\nu}^{K}}\partial_{\tau}^{K}\alpha_{K}
\label{33}  
\end{equation} 
with a Newton constant $m_{\nu}^{K}$ at the level of inertial mass of any $(K)$ nature. The quantum operator $\partial_{\tau}^{K}$ with Newton behavior may therefore be written as
\begin{equation}
\partial_{\tau}^{K}=\partial_{\tau}^{K}-m_{\nu}^{K}A_{\tau}^{K}.
\label{34}  
\end{equation} 

Thus, using (\ref{23}) and (\ref{34}) and having in mind (\ref{31}) and (\ref{32}), one can find from (\ref{29}) that the Lagrangian $L^{D}$ invariance concerning each pair of local gauge transformations (\ref{21}), (\ref{22}), (\ref{30}), and (\ref{33}) includes a kind of Newton part and a kind of Coulomb part of the naturally united gravitational interaction
$$L^{D}=L_{free}^{D}+L_{int}^{D}=$$
\begin{equation}
=i\overline{\psi}_{\nu}(\gamma^{\mu}\partial_{\mu}^{s}+\partial_{\tau}^{s})\psi_{\nu}-
ie_{\nu}^{K}j^{\mu}_{C}A_{\mu}^{K}-im_{\nu}^{K}j^{\tau}_{N}A_{\tau}^{K}.
\label{35}  
\end{equation} 

In it appears a part of the dynamical origin of mass $m_{\nu}^{K}$ and charge $e_{\nu}^{K}$ 
in the unified mass-charge structure dependence of gauge invariance. Objects with $m_{\nu}^{K}$ 
and $e_{\nu}^{K}$ are of course the source of Newton $j^{\tau}_{N}$ and Coulomb $j^{\mu}_{C}$ components, respectively, of a C-invariant graviton current of any $(K)$ nature
\begin{equation}
j^{\mu}_{C}=\overline{\psi}_{\nu}\gamma^{\mu}\psi_{\nu}, \, \, \, \,
j^{\tau}_{N}=\overline{\psi}_{\nu}\psi_{\nu}.
\label{36}  
\end{equation} 

However, as noted in (\ref{35}), the field $A_{\tau}^{K}$ equalized with the Newton field is not of the fields such that their source may serve a single object with an electric mass. In other words, the appearance of any Newton field $A_{\tau}^{K}$ in a Lagrangian (\ref{35}) at $K=E,$ $W,$ $S,...$ 
explains the availability in each type of interaction of a kind of mediating boson of a Newton nature. It is one of those gauge bosons, each of which gives a kind of gravitational mass 
to all objects that interact with its matter field.

This is exactly the same as when $K\rightarrow H_{K},$ $N\rightarrow H_{N},$ and 
$\nu\rightarrow \nu_{H}.$ Consequently, one of the magnetoelectric $(H_{E}),$ magnetoweak $(H_{W}),$ and magnetostrong $(H_{S})$ parts and the range of other innate parts of the magnetogravitational $(H_{G})$ mass of an elementary monoparticle arises at the expense of a kind of interaction between this monoparticle and one of the gauge monobosons $(A_{\tau}^{H_{K}})$ having a magnetonewton $(H_{N})$ nature.

The masses of a mononeutrino $\nu_{H}$ hereby constitute the united $(U_{H})$ magnetogravitational $(H_{G})$ mass equal to its grand magnetounited $(GU_{H})$ mass
\begin{equation}
m_{\nu_{H}}=m_{\nu_{H}}^{U_{H}}=m_{\nu_{H}}^{H_{G}}=m_{\nu_{H}}^{GU_{H}}=
m_{\nu_{H}}^{H_{E}}+m_{\nu_{H}}^{H_{W}}+m_{\nu_{H}}^{H_{S}}+...
\label{37}
\end{equation}
expressing the magnetonewton forces among two mononeutrinos as 
\begin{equation}
F^{H_{K}}_{H_{N}^{\nu_{H}\nu_{H}}}=
G_{H_{N}}\left(\frac{m_{\nu_{H}}^{H_{K}}}{r_{H}}\right)^{2}, \, \, \, \,
F^{H_{i}H_{j}}_{H_{N}^{\nu_{H}\nu_{H}}}=
G_{H_{N}}\frac{m_{\nu_{H}}^{H_{i}}m_{\nu_{H}}^{H_{j}}}{r_{H}^{2}},
\label{38}
\end{equation}
where $G_{H_{N}}$ denotes the constant of magnetogravitation.

\vspace{0.6cm}
\noindent
{\bf 8. Dynamical origin of mass and charge
\\of gauge bosons and monobosons}
\vspace{0.3cm}

An intragraviton feature of the two types of bosons of any $(K)$ nature is their coexistence. As 
a consequence, each of the Coulomb $(\gamma_{K}^{C})$ mediating bosons testifies in favor of the existence of a kind of Newton $(\gamma_{K}^{N})$ mediating boson. They hereby constitute the 
united Newton-Coulomb pairs of gauge bosons. 

An important characteristic of a general picture of these gauge pairs is that the free boson Lagrangian with a choice of the four-component wave function of each structural particle $\varphi_{\gamma_{K}}(t, {\bf x})$ and its mass $m_{\gamma_{K}}^{K}=-i\partial_{\tau}^{K}$ 
may have the form [21]
\begin{equation}
L_{free}^{B}=\frac{1}{2}\varphi_{\gamma_{K}}^{*}(\partial_{\mu}^{K}\partial^{\mu}_{K}-\partial_{\tau}^{K}\partial^{\tau}_{K})\varphi_{\gamma_{K}}
\label{39}  
\end{equation} 
in which it is definitely stated that
\begin{equation}
\partial_{\mu}^{K}\varphi_{\gamma_{K}}=\partial_{\mu}^{K}\varphi_{\gamma_{K}}(x), \, \, \, \, 
\partial_{\tau}^{K}\varphi_{\gamma_{K}}=\partial_{\tau}^{K}\varphi_{\gamma_{K}}(\tau).
\label{40}
\end{equation}  

Of course, the boson Lagrangian (\ref{39}) is noninvariant for the local phases $\alpha_{K}(x)$ and 
$\alpha_{K}(\tau)$ of gauge transformations such as 
\begin{equation}
\varphi'_{\gamma_{K}}=U_{K}^{C}\varphi_{\gamma_{K}}, \, \, \, \, U_{K}^{C}=e^{i\alpha_{K}(x)},
\label{41}  
\end{equation} 
\begin{equation}
\varphi'_{\gamma_{K}}=U_{K}^{N}\varphi_{\gamma_{K}}, \, \, \, \, U_{K}^{N}=e^{i\alpha_{K}(\tau)}.
\label{42}  
\end{equation} 

To restore its broken internal symmetry, it is desirable to introduce a Newton-Coulomb pair of gauge fields $A_{\mu}^{K}$ and $A_{\tau}^{K}$ with the corresponding pairs of gauge transformations. 

Therefore, if one uses the substitutions
\begin{equation}
e_{\nu}^{K}\rightarrow e_{\gamma_{K}}^{K}, \, \, \, \, 
m_{\nu}^{K}\rightarrow m_{\gamma_{K}}^{K}
\label{43}  
\end{equation} 
and then performs explicit gauge transformations, one can establish an invariance of 
the boson Lagrangian 
$$L^{B}=L_{free}^{B}+L_{int}^{B}=$$
$$=\frac{1}{2}\varphi_{\gamma_{K}}^{*}(\partial_{\mu}^{K}\partial^{\mu}_{K}-
\partial_{\tau}^{K}\partial^{\tau}_{K})\varphi_{\gamma_{K}}+$$
$$+\frac{1}{2}[e_{\gamma_{K}}^{K}(J_{\mu}^{C}A^{\mu}_{K}-J^{\mu}_{C}A_{\mu}^{K})-
(e_{\gamma_{K}}^{K})^{2}\varphi_{\gamma_{K}}^{*}\varphi_{\gamma_{K}}A_{\mu}^{K}A^{\mu}_{K}]-$$
\begin{equation}
-\frac{1}{2}[m_{\gamma_{K}}^{K}(J_{\tau}^{N}A^{\tau}_{K}-J^{\tau}_{N}A_{\tau}^{K})-
(m_{\gamma_{K}}^{K})^{2}\varphi_{\gamma_{K}}^{*}\varphi_{\gamma_{K}}A_{\tau}^{K}A^{\tau}_{K}]
\label{44}  
\end{equation} 
concerning the action of selected pairs of the local gauge transformations (\ref{22}), (\ref{33}), (\ref{41}), and (\ref{42}) due to an interaction of all types of bosons with the unified 
Newton-Coulomb field of the graviton of any $(K)$ nature.

So it is seen that gauge fields $A_{\mu}^{K}(A^{\mu}_{K})$ and $A_{\tau}^{K}(A^{\tau}_{K})$
interact, respectively, with the Coulomb $J^{\mu}_{C}(J_{\mu}^{C})$ and Newton 
$J^{\tau}_{N}(J_{\tau}^{N})$ currents
\begin{equation}
J^{\mu}_{C}=\varphi_{\gamma_{K}}^{*}\partial^{\mu}_{K}\varphi_{\gamma_{K}}, \, \, \, \,
J_{\mu}^{C}=\varphi_{\gamma_{K}}^{*}\partial_{\mu}^{K}\varphi_{\gamma_{K}},
\label{45}  
\end{equation} 
\begin{equation}
J^{\tau}_{N}=\varphi_{\gamma_{K}}^{*}\partial^{\tau}_{K}\varphi_{\gamma_{K}}, \, \, \, \,
J_{\tau}^{N}=\varphi_{\gamma_{K}}^{*}\partial_{\tau}^{K}\varphi_{\gamma_{K}}
\label{46}  
\end{equation} 
forming the two components of the same boson current. 

From their point of view, each of the electric $(\gamma_{E}),$ weak $(\gamma_{W}),$ and strong $(\gamma_{S})$ gauge bosons and the range of other innate types of gauge bosons may have a latently united structure
\begin{equation}
\gamma_{E}=\{\gamma_{E}^{C}, \, \, \, \, \gamma_{E}^{N}\}, 
\label{47}
\end{equation}
\begin{equation}
\gamma_{W}=\{\gamma_{W}^{C}, \, \, \, \, \gamma_{W}^{N}\}, 
\label{48}
\end{equation} 
\begin{equation}
\gamma_{S}=\{\gamma_{S}^{C}, \, \, \, \, \gamma_{S}^{N}\}, \, \, \, \, ...
\label{49}
\end{equation}
at the new level.

These united states convince us that 
\begin{equation}
\gamma_{GU}^{C}=
\{\gamma_{E}^{C}, \, \, \, \, \gamma_{W}^{C}, \, \, \, \, \gamma_{S}^{C}, \, \, \, \, ...\},
\label{50}
\end{equation}
\begin{equation}
\gamma_{GU}^{N}=
\{\gamma_{E}^{N}, \, \, \, \, \gamma_{W}^{N}, \, \, \, \, \gamma_{S}^{N}, \, \, \, \, ...\}.
\label{51}
\end{equation}

Thus, between Newton and Coulomb gravitons, there exists a range of structural connections 
that are united in a single graviton
\begin{equation}
\gamma_{GU}=\{\gamma_{GU}^{C}, \, \, \, \, \gamma_{GU}^{N}\}
\label{52}
\end{equation}
such that it gives a kind of gravitational mass and a kind of gravitational charge to all objects that interact with its matter fields and comes forward in nature as the united gauge boson, namely, as the mediating graviton of the naturally united gravitational $(GU)$ interaction.

However, we have mentioned that the magnetogravitational $(H_{G})$ mass and charge of each of the magnetoelectric $(H_{E}),$ magnetoweak $(H_{W}),$ and magnetostrong $(H_{S})$ gauge monobosons and the range of other innate types of gauge monobosons 
\begin{equation}
\gamma_{H_{E}}=\{\gamma_{H_{E}}^{H_{C}}, \, \, \, \, \gamma_{H_{E}}^{H_{N}}\},
\label{53}
\end{equation}
\begin{equation}
\gamma_{H_{W}}=\{\gamma_{H_{W}}^{H_{C}}, \, \, \, \, \gamma_{H_{W}}^{H_{N}}\},
\label{54}
\end{equation}
\begin{equation}
\gamma_{H_{S}}=\{\gamma_{H_{S}}^{H_{C}}, \, \, \, \, \gamma_{H_{S}}^{H_{N}}\}, \, \, \, \, ...
\label{55}
\end{equation}
are the consequences of the connection that it, similarly to all other massive monomatter, has undergone a fully latent interaction with a monograviton
\begin{equation}
\gamma_{GU_{H}}=\{\gamma_{GU_{H}}^{H_{C}}, \, \, \, \, \gamma_{GU_{H}}^{H_{N}}\}.
\label{56}
\end{equation}

Magnetonewton and magnetocoulomb monogravitons
\begin{equation}
\gamma_{GU_{H}}^{H_{C}}=
\{\gamma_{H_{E}}^{H_{C}}, \, \, \, \, \gamma_{H_{W}}^{H_{C}}, \, \, \, \,
\gamma_{H_{S}}^{H_{C}}, \, \, \, \, ...\},
\label{57}
\end{equation}
\begin{equation}
\gamma_{GU_{H}}^{H_{N}}=
\{\gamma_{H_{E}}^{H_{N}}, \, \, \, \, \gamma_{H_{W}}^{H_{N}}, \, \, \, \,
\gamma_{H_{S}}^{H_{N}}, \, \, \, \, ...\}
\label{58}
\end{equation}
are united thus in a unified whole as the magnetounited gauge monoboson, namely, as the mediating monograviton of the naturally united magnetogravitational $(GU_{H})$ interaction.

If two mononeutrinos interact, one can define the structure of magnetonewton and magnetocoulomb forces for each pair of naturally magnetounited $(GU_{H})$ masses and charges:
\begin{equation}
F^{GU_{H}}_{H_{N}^{\nu_{H}\nu_{H}}}=
G_{H_{N}}\left(\frac{m_{\nu_{H}}^{GU_{H}}}{r_{H}}\right)^{2}, \, \, \, \,
F^{GU_{H}}_{H_{C}^{\nu_{H}\nu_{H}}}=\frac{1}{4\pi\epsilon_{0}}
\left(\frac{e_{\nu_{H}}^{GU_{H}}}{r_{H}}\right)^{2}.
\label{59}
\end{equation}

The insertion of (\ref{25}) and (\ref{37}) into (\ref{59}) and unification of the findings 
with (\ref{26}) and (\ref{38}) at
\begin{equation}
F^{H_{K}}_{\nu_{H}\nu_{H}}=
F^{H_{K}}_{H_{N}^{\nu_{H}\nu_{H}}}+F^{H_{K}}_{H_{C}^{\nu_{H}\nu_{H}}},
\label{60}
\end{equation}
\begin{equation}
F^{H_{i}H_{j}}_{\nu_{H}\nu_{H}}=
F^{H_{i}H_{j}}_{H_{N}^{\nu_{H}\nu_{H}}}+F^{H_{i}H_{j}}_{H_{C}^{\nu_{H}\nu_{H}}}
\label{61}
\end{equation}
constitute a system 
\begin{equation}
F^{GU_{H}}_{H_{N}^{\nu_{H}\nu_{H}}}=
F^{H_{E}}_{H_{N}^{\nu_{H}\nu_{H}}}+F^{H_{W}}_{H_{N}^{\nu_{H}\nu_{H}}}+
F^{H_{S}}_{H_{N}^{\nu_{H}\nu_{H}}}+...,
\label{62}
\end{equation}
\begin{equation}
F^{GU_{H}}_{H_{C}^{\nu_{H}\nu_{H}}}=
F^{H_{E}}_{H_{C}^{\nu_{H}\nu_{H}}}+F^{H_{W}}_{H_{C}^{\nu_{H}\nu_{H}}}+
F^{H_{S}}_{H_{C}^{\nu_{H}\nu_{H}}}+....
\label{63}
\end{equation}

These equalities and relationship 
\begin{equation}
F^{GU_{H}}_{\nu_{H}\nu_{H}}=
F^{GU_{H}}_{H_{N}^{\nu_{H}\nu_{H}}}+F^{GU_{H}}_{H_{C}^{\nu_{H}\nu_{H}}}
\label{64}
\end{equation}
state that 
\begin{equation}
F^{GU_{H}}_{\nu_{H}\nu_{H}}=
F^{H_{E}}_{\nu_{H}\nu_{H}}+F^{H_{W}}_{\nu_{H}\nu_{H}}+ F^{H_{S}}_{\nu_{H}\nu_{H}}+...,
\label{65}
\end{equation}
which leads to the appearance of naturally united gauge fields of monomatter. This has become possible because of the unified mass-charge structure of all types of magnetic forces. They are, 
of course, the structural components of the interaction magnetounited force, which comes forward 
in nature as magnetogravity.

For completeness, we recall the photon-monophoton duality [15], according to which an electric boson of a graviton says about the existence in a monograviton of a kind of magnetoelectric monoboson. Therefore, without loss of the unity of symmetry laws, each intragraviton boson corresponds to a kind of intramonograviton monoboson. Thus, nature unites the graviton and monograviton in a unified whole as a grand united gauge allboson, namely, as the mediating allgraviton
\begin{equation}
\gamma_{AG}=\{\gamma_{GU}, \, \, \, \, \gamma_{GU_{H}}\}
\label{66}
\end{equation}
of the naturally united allgravitational $(AG)$ interaction. An allgravity hereby comes forward  
as a grand unification of forces. 

\vspace{0.6cm}
\noindent
{\bf 9. An allgraviton about the gauge group of a grand unification}
\vspace{0.3cm}

Passing to the question about the gauge group of the suggested grand unification, one can first 
recall that the standard electroweak theory is based [28-30] on groups such as $SU(2)_{L}$ and $U(1)_{\gamma},$ which constitute an electroweakly united group $G_{EW}$ in the form
\begin{equation}
G_{EW}\supset SU(2)_{L}\otimes U(1)_{\gamma}.
\label{67}
\end{equation}

In this theory, the right-handed neutrinos $\nu_{R} ({\bar \nu_{L}})$ have 
neither weak, electromagnetic interaction nor any other interaction, and the left-handed neutrinos 
$\nu_{L} ({\bar \nu_{R}})$ can interact only with the field of weak emission. In addition, it is accepted that the same photon leads to the appearance of electric and magnetic fields.

A given circumstance seems to require one to follow the coexistence law [33] of each pair of a 
lepton and its neutrino because it constitutes the leptonic families of the left-handed $SU(2)_{L}$ and right-handed $SU(2)_{R}$ doublets. Their existence extends the group (\ref{67}) 
in the following manner:
\begin{equation}
G_{EW} \supset SU(2)_{L}\otimes SU(2)_{R}\otimes U(1)_{\gamma}.
\label{68}
\end{equation}

Therefore, to extend group $G_{EW},$ we must recall the compound structure of an electromagnetic boson (\ref{17}) that
\begin{equation}
U(1)_{\gamma}\rightarrow     
U(1)_{\gamma_{EH_{E}}}\Rightarrow U(1)_{\gamma_{E}}\otimes U(1)_{\gamma_{H_{E}}}.
\label{69}
\end{equation}

In these circumstances, the first-initial presentation (\ref{67}) is replaced for
\begin{equation}
G_{EW} \supset SU(2)_{L}\otimes SU(2)_{R}\otimes U(1)_{\gamma_{E}}\otimes
U(1)_{\gamma_{H_{E}}},
\label{70}
\end{equation}
and consequently, in the framework of such an extended electroweak theory, there exists a hard connection between left and right and between electricity and magnetism.

Furthermore, if the lepton number could be accepted as the fourth color [34], this would indicate 
the existence of color $(C)$ group $SU(4)_{C},$ which involves all neutrinos having a nonzero 
strong interaction [35]. The latter together with a group (\ref{70}) implies that
\begin{equation}
G_{GU} \supset SU(4)_{C}\otimes SU(2)_{L}\otimes
SU(2)_{R}\otimes U(1)_{\gamma_{E}}\otimes U(1)_{\gamma_{H_{E}}}.
\label{71}
\end{equation}

These connections seem to indicate that the gauge group of a grand unification at the level predicted by electroweak theory describes nature without loss of generality. On the other hand, as is known, the standard electroweak theory, by itself, does not give a mass to any of the existing types of matter fields and requires their interaction with Higgs [36] bosons, the existence of which would imply its trueness. However, according to a mass-charge structure of gauge invariance, this would take place only in the case of the unity of symmetry laws being wholly absent. Therefore, without contradicting ideas of the local gauge transformations (\ref{21}) and (\ref{30}), the abelian groups $U_{K}^{C}(1)$ and $U_{K}^{N}(1)$ can in the limits of fermion (boson) and graviton fields constitute a single abelian gauge group 
\begin{equation}
G_{GU} \supset U_{GU}^{C}(1)\otimes U_{GU}^{N}(1)\otimes
U_{GU}^{C}(1)_{\gamma_{GU}}\otimes U_{GU}^{N}(1)_{\gamma_{GU}}
\label{72}
\end{equation}
such that it describes gravity.

This, in turn, implies that the magnetoabelian groups $U_{H_{K}}^{H_{C}}(1)$ and 
$U_{H_{K}}^{H_{N}}(1)$ can in the limits of monofermion (monoboson) and monograviton 
fields constitute a single magnetoabelian gauge group 
\begin{equation}
G_{GU_{H}} \supset U_{GU_{H}}^{H_{C}}(1)\otimes U_{GU_{H}}^{H_{N}}(1)\otimes
U_{GU_{H}}^{H_{C}}(1)_{\gamma_{GU_{H}}}\otimes U_{GU_{H}}^{H_{N}}(1)_{\gamma_{GU_{H}}}
\label{73}
\end{equation}
such that it describes a magnetogravity.

Thus, our presentations of the true nature of symmetry between electricity and magnetism allow 
not only the use of this symmetrical structure at the level of a grand unification mathematical 
logic as the defined symmetry between gravity and magnetogravity but also their unification in 
a single allgravity $(AG)$ responsible for all that in a curved space-time of allgraviton fields. 
The abelian and magnetoabelian groups hereby constitute a single allabelian gauge group 
\begin{equation}
G_{AG} \supset G_{GU}\otimes G_{GU_{H}} 
\label{74}
\end{equation}
such that it describes an allgravity.

\vspace{0.6cm}
\noindent
{\bf 10. Ether from the point of view of gravity}
\vspace{0.3cm}

If we now consider that no one is in force to separate any particle by part in the mass or charge type dependence, the gravitational field undoubtedly comes forward in nature as a unified whole. Therefore, according to the suggested theory of grand unification, one form of ether should be expected to exist as a gravity, namely, as a unification of the naturally united gauge fields.

To characterize such an unusual picture from the point of view of the well-known Michelson-Morley experiment [37-39], it is very important to elucidate whether there exists any connection between 
the mass of a photon and its spin nature and, if so, what the expected dependence says about the scattering of light on the suggested world medium.

For this purpose, we must first recall the massless neutrino [40]. It is strictly longitudinally polarized. However, under the availability of a nonzero mass, the longitudinal neutrino in the nucleus charge field can be converted into a transverse one and vice versa [27]. In other words, the same neutrino of nonzero mass must have either longitudinal or transverse polarization. It is fully possible, therefore, that transitions between two neutrinos of different polarizations indicate the existence of fundamental differences in nature of the masses of longitudinal
and transverse particles [27].

There exists a range of other phenomena in which appears a connection between photons of different polarizations. The same massive photon must not be simultaneously a longitudinal and transverse boson. It cannot have neither the longitudinal nor transverse polarization if it has no mass. Unlike neutrinos, photons with zero mass are strictly circularly polarized. Then, it is possible, for example, that the mass in a photon is available such that it transforms its circular polarization into longitudinal or transverse polarization. This, however, does not exclude that a longitudinal photon in the nucleus field can be converted into a transverse one and vice versa.

A given circumstance may serve as a certain indication of an explicit spin polarization type dependence of the photon mass and nature.

Thus, it follows that the difference in nature of longitudinal and transverse particles leads to their interconversion. In these conditions, an incoming flux of longitudinal (transverse) polarized neutrinos or photons suffers considerable warping of the trajectory at their passage through 
the nucleus.

From the point of view of the nuclear target itself, a similar angular deflection testifies 
in favor of its periodic revolution around its axis, at which the curvature of the field of an 
action is strongly changed. An analogous situation takes place in the case where an instantaneous reestablishment of the harmony of the interacting forces originates. Such an order, however, can exist regardless of the source of the field.

Returning to the Michelson-Morley ether experiment [37-39], we now recall that its purpose was to discover the absolute speed of an ether wind relative to the Earth and thus to directly establish 
the existence of a truly stationary frame of reference in which it is at rest.

To this end, a physical instrument (Fig. 1) called the Michelson interferometer was used. An 
incoming light beam $A$ in this instrument with the aid of a half-silvered glass plate is separated into two interperpendicular parts $C$ and $D$ traveling at the corresponding speeds relative to the ether. Next, these beams are reflected from the non-transparent mirrors placed at equal distances from a half-transparent glass plate. The optical system is such that beams $C$ and $D$ are returned to the same screen in which appears their interference picture.

\begin{center}
{\thicklines
\begin{picture}(120,100)
\put(100,25){\rule{5\unitlength}{35\unitlength}}
\put(50,80){\rule{35\unitlength}{5\unitlength}}
\put(50,35.05){\line(1,1){30}}\put(80,65.05){\line(1,-1){5}}
\put(55,30.05){\line(1,1){30}}\put(50,35.05){\line(1,-1){5}}
{\thicklines\put(15,50){\vector(1,0){35}}
\put(15,45){\vector(1,0){35}} \put(65,50){\vector(0,1){30}}
\put(70,80){\vector(0,-1){15}} \put(50,45){\vector(1,0){50}}
\put(100,40){\vector(-1,0){15}} \put(65,40){\vector(0,-1){20}}
\put(70,65){\vector(0,-1){40}}
\put(15,70){\vector(1,0){10}}\put(110,70){\vector(1,0){10}}
\put(15,15){\vector(1,0){10}}\put(110,15){\vector(1,0){10}}
\put(50,50){\line(1,0){15}}\put(85,40){\line(-1,0){20}}
\put(65,20){\line(0,-1){10}}\put(70,25){\line(0,-1){15}}}
\put(18,72){$E$}\put(113,72){$E$}\put(18,17){$E$}\put(112,17){$E$}
\put(10,46){$A$}\put(59,73){$C$}\put(93,47){$D$}
\put(72,23){$C$}\put(59,16){$D$}
\put(62,87){mirror}\put(107,41){mirror}
\put(15,52){incoming light}\put(23,30){beam splitter}\put(60,5){detector}
\end{picture}}\\[0.0cm]
\end{center}
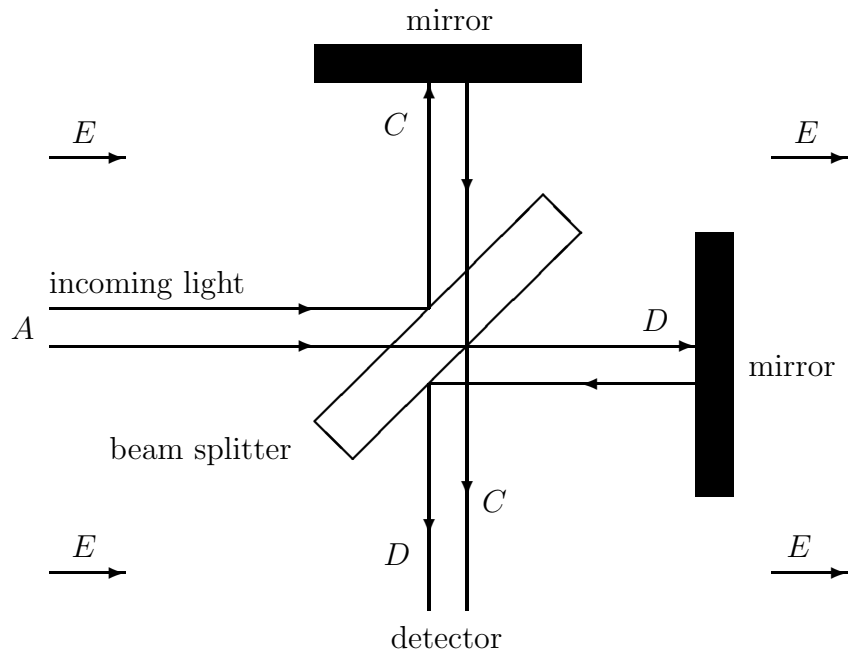
Fig. 1. Scheme of the Michelson-Morley ether $(E)$ experiment.
\vspace{0.8cm}

If any of the light beams $C$ or $D$ in an interferometer changes the length of its path, the picture 
of interference bands must be moved in the optical device. To observe this shift, one must rotate the instrument around the stationary basis at an angle of $90^{0},$ at which the two light beams are exchanged by the orientations of their paths (Fig. 2) relative to the direction of the ether wind. 
In these circumstances, in Michelson and Morley's opinion, one of the light beams $C(D)$ or $C'(D')$ at its passage through the ether from the half-silvered glass plate to the detector suffers a delay in time. As a consequence, the interference picture will move in the optical system in the apparatus rotational dependence.

According to the description of both authors, the interference bands of the two observed beams in comparison with a certain middle position of the screen twice for one turn of an instrument must 
move to the right direction and then to the left one and vice versa.

\begin{center}
{\thicklines
\begin{picture}(120,100)
\put(100,25){\rule{5\unitlength}{35\unitlength}}
\put(50,80){\rule{35\unitlength}{5\unitlength}}
\put(50,35.05){\line(1,1){30}}\put(80,65.05){\line(1,-1){5}}
\put(55,30.05){\line(1,1){30}}\put(50,35.05){\line(1,-1){5}}
{\thicklines\put(15,50){\vector(1,0){35}}
\put(15,45){\vector(1,0){35}} \put(65,50){\vector(0,1){30}}
\put(70,80){\vector(0,-1){15}} \put(50,45){\vector(1,0){50}}
\put(100,40){\vector(-1,0){15}} \put(65,40){\vector(0,-1){20}}
\put(70,65){\vector(0,-1){40}}
\put(20,65){\vector(0,1){10}}\put(115,65){\vector(0,1){10}}
\put(20,10){\vector(0,1){10}}\put(115,10){\vector(0,1){10}}
\put(50,50){\line(1,0){15}}\put(85,40){\line(-1,0){20}}
\put(65,20){\line(0,-1){10}}\put(70,25){\line(0,-1){15}}}
\put(15,67){$E$}\put(110,67){$E$}\put(15,12){$E$}\put(110,12){$E$}
\put(10,46){$A$}\put(59,73){$C'$}\put(93,47){$D'$}
\put(72,23){$C'$}\put(59,16){$D'$}
\put(62,87){mirror}\put(107,41){mirror}
\put(15,52){incoming light}\put(23,30){beam splitter}\put(60,5){detector}
\end{picture}}\\[0.0cm]
\end{center}
Fig. 2. Ether $(E)$ under the rotation of a Michelson interferometer.
\vspace{0.8cm}

This shift has not been detected. However, unlike the earlier expectation, the displacement of interference pictures depending on the angle of turn of the device was only in one direction, 
namely, to the right or to the left. 

Under these conditions, proof has been obtained that ether does not exist, and the speed of 
light in a vacuum is the same in all directions and does not depend on the motion of its source 
and the observer.

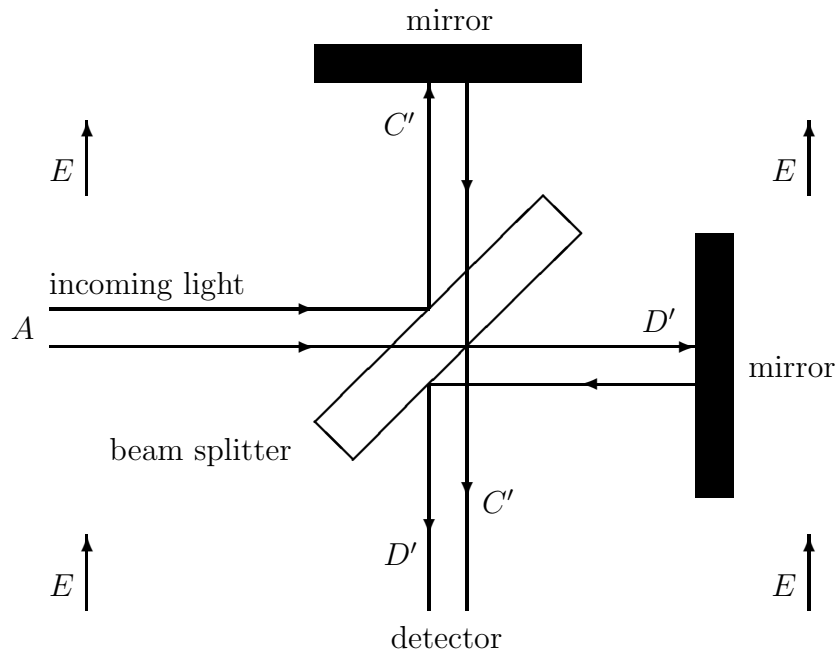
Of course, such an implication was based on the assumption that the existence of an ether sea would lead to a time delay of one of the light beams $C(D)$ or $C'(D')$ at the rotation of the instrument. This is explained by predictions of the classical theory of ether wind. It states that ether is strictly truly a stationary medium in which all physical objects move, and therefore, their 
velocity is intimately connected with its nature.

The absence of an ether sea implies the availability of all matter in absolute emptiness. However, this possibility would be realized if and only if empty space, through which all matter has passed, comes forward in the universe as a single space, confirming that it is not curved.

Our previous analyses show that the unified world picture of all physical phenomena has been created on the basis of a grand unification of forces. Their field is strictly curved. This curvature leads us to the conclusion that one of the forms of ether exists as gravity, latently acting on all matter without exception. 

Thus, the purpose of the Michelson-Morley experiment is reduced to the definition of the speed 
of a gravitational wind relative to the Earth and thereby to the establishment of its existence.

As a consequence, at the turn of the interferometer at an angle of $90^{0},$ the two light paths are exchanged by their orientations in comparison with the direction of a gravitational wind. In these conditions, one of the light beams $C(D)$ or $C'(D')$ passing through the gravitational sea from 
the half-transparent glass plate to the optical device, similarly to the abovementioned angular  deflection in the nucleus field of the flux of longitudinal (transverse) polarized neutrinos or photons, suffers considerable warping of its trajectory, and is not delayed. Therefore, the observed shift of interference bands depending on the angle of rotation of the apparatus must be either to the right or left direction relative to a certain middle position.

Such a geometrical picture of the phenomenon discovered in the Michelson-Morley experiment may 
serve as the first laboratory confirmation of an angular deflection of light as a consequence of the interconversion of light beams in their spin polarization type dependence. However, from the point 
of view of a general theory of relativity, gravity can essentially change the trajectory of light. Thus, if absolute emptiness itself is not in force to bend light regardless of gravity, we must recognize that one of the forms of an ether sea exists as a gravitational field.

\vspace{0.6cm}
\noindent
{\bf 11. Light-force duality principle}
\vspace{0.3cm}

To better express the idea, it is desirable to refer again to the light in a Michelson interferometer because it has an electric structure. Its source was a unified system of the Coulomb and Newton types of photons, which comes forward in nature as either an electric boson (\ref{47}) or an electric wave.

Its structural bosons suffer periodic interconversion in which the Coulomb force is 
converted to a Newton one and vice versa. The speed of these transitions coincides with the individual velocities of both types of photons and does not depend on the speed of the system. Furthermore, electric light with its own speed has longitudinal or transverse spin polarization, because of which an ether, namely, gravity, bends its trajectory in a Michelson interferometer. 

Such a presentation is logically based on the unified principle that each gauge boson may serve as the source of a kind of light. The existence of weak and strong light beams and the range of other innate types of light beams is hereby not excluded. They together with an electric light constitute 
a gravitational light, which comes forward in the universe as either a flux of gravitons (\ref{52}) 
or a gravitational wave. 

To investigate further, we can use, for example, in a Michelson interferometer, another light 
such that it has a magnetoelectric nature. Its source is a unified system of the magnetocoulomb 
and magnetonewton types of monophotons. It comes forward in the universe as either 
a magnetoelectric monoboson (\ref{53}) or a magnetoelectric wave.

The monobosons suffer periodic interconversion within a monophoton in which the magnetocoulomb 
force is converted to a magnetonewton force and vice versa. Their speed coincides with the velocities 
of both types of monophotons and does not depend on the speed of the magnetosystem itself. At the same time, magnetoelectric light itself with an individual velocity possesses either a longitudinal 
or transverse spin polarization, because of which a magnetoether, namely, a magnetogravity, bends 
its trajectory in a Michelson interferometer. 

Each gauge monoboson may serve as the source of a kind of light. Such pairs testify in favor of  magnetoweak and magnetostrong light beams and the range of other innate types of light beams. With 
a magnetoelectric light, they constitute magnetogravitational light, which comes forward in nature 
as either a flux of monogravitons (\ref{56}) or a magnetogravitational wave. 

It is here that we must, for the first time, use the northern lights of the natural origin, emphasizing that it was always the northern magnetogravitational lights.

Natural light, namely, the light of astronomical objects, has hereby a gravitational and
magnetogravitational nature. Its source may serve a unified system of graviton and monograviton. 
They constitute the structural components of an allgravitational light. It unites the gravitational and magnetogravitational lights in a unified whole as natural light.

Thus, unlike the earlier presentations on the unity of matter fields, the discussed theory of grand unification leads us to a correspondence principle that each type of force testifies in favor of the existence of a kind of light. Their nature has been created so that any type of light corresponds to a kind of force. If a given situation follows from a unified principle, the light and force correspond to two forms of the same matter. Such a correspondence principle expresses [41] 
the light-force duality. 

\vspace{0.6cm}
\noindent
{\bf 12. Conclusion}
\vspace{0.3cm}

According to the suggested theory of grand unification, an allgravity is responsible for a generalized principle of Einstein relativity because it unites gravity and magnetogravity. 
In other words, the ether and magnetoether correspond to two forms of the same allether. Of course, all physical phenomena originate in an allether sea, namely, in an absolutely curved allgravitational sea. No one is hereby in force to observe the strictly straight linear and uniform motion of one particle (monoparticle) relative to a second one. Such an world curvature reflects the harmony 
of the naturally curved forces.

Another important implication of the northern magnetogravitational lights is that there 
is no single pair of Coulomb (Newton) and magnetocoulomb (magnetonewton) fields of an allgravity, 
for which a single pair from $K$ and $H_{K}$ systems of wave equations of grand unification would 
not exist analogously to the fact that for one pair of electric and magnetic fields of classical electrodynamics there exists one selected pair from the systems of well-known Maxwell equations. 
However, we can establish the mathematical connections of allcoulomb  $\vec{\gamma}_{KH_{K}}^{C}$ and allnewton $\vec{\gamma}_{KH_{K}}^{N}$ fields within each of the allelectric, allweak, and allstrong pairs and the range of other innate types of pairs 
\begin{equation}
\vec{\gamma}_{EH_{E}}^{C}=
\{\vec{\gamma}_{E}^{C}, \, \, \, \, \vec{\gamma}_{H_{E}}^{H_{C}}\}, \, \, \, \,
\vec{\gamma}_{WH_{W}}^{C}=
\{\vec{\gamma}_{W}^{C}, \, \, \, \, \vec{\gamma}_{H_{W}}^{H_{C}}\}, \, \, \, \,
\vec{\gamma}_{SH_{S}}^{C}=
\{\vec{\gamma}_{S}^{C}, \, \, \, \, \vec{\gamma}_{H_{S}}^{H_{C}}\}, \, \, \, \, ...,
\label{75}
\end{equation}
\begin{equation}
\vec{\gamma}_{EH_{E}}^{N}=
\{\vec{\gamma}_{E}^{N}, \, \, \, \, \vec{\gamma}_{H_{E}}^{H_{N}}\}, \, \, \, \,
\vec{\gamma}_{WH_{W}}^{N}=
\{\vec{\gamma}_{W}^{N}, \, \, \, \, \vec{\gamma}_{H_{W}}^{H_{N}}\}, \, \, \, \,
\vec{\gamma}_{SH_{S}}^{N}=
\{\vec{\gamma}_{S}^{N}, \, \, \, \, \vec{\gamma}_{H_{S}}^{H_{N}}\}, \, \, \, \, ...
\label{76}
\end{equation}
if and only if the very space in which it describes one of the allcoulomb $\gamma_{KH_{K}}^{C}$ 
and allnewton $\gamma_{KH_{K}}^{N}$ mediating bosons 
\begin{equation}
\gamma_{EH_{E}}^{C}=\{\gamma_{E}^{C}, \, \, \, \, \gamma_{H_{E}}^{H_{C}}\}, \, \, \, \,
\gamma_{WH_{W}}^{C}=\{\gamma_{W}^{C}, \, \, \, \, \gamma_{H_{W}}^{H_{C}}\}, \, \, \, \,
\gamma_{SH_{S}}^{C}=\{\gamma_{S}^{C}, \, \, \, \, \gamma_{H_{S}}^{H_{C}}\}, \, \, \, \, ...,
\label{77}
\end{equation}
\begin{equation}
\gamma_{EH_{E}}^{N}=\{\gamma_{E}^{N}, \, \, \, \, \gamma_{H_{E}}^{H_{N}}\}, \, \, \, \,
\gamma_{WH_{W}}^{N}=\{\gamma_{W}^{N}, \, \, \, \, \gamma_{H_{W}}^{H_{N}}\}, \, \, \, \,
\gamma_{SH_{S}}^{N}=\{\gamma_{S}^{N}, \, \, \, \, \gamma_{H_{S}}^{H_{N}}\}, \, \, \, \, ...
\label{78}
\end{equation}
unites all structural fields in the two classes. To the first of them apply the curved vector 
fields $\vec{\gamma}_{K}^{C} (\vec{\gamma}_{K}^{N})$ of bosons $\gamma_{K}^{C} (\gamma_{K}^{N}),$ which has no magnetocoulomb (magnetonewton) nature. The second class consists of the curved vector 
fields $\vec{\gamma}_{H_{K}}^{H_{C}} (\vec{\gamma}_{H_{K}}^{H_{N}})$ of monobosons 
$\gamma_{H_{K}}^{H_{C}} (\gamma_{H_{K}}^{H_{N}}),$ in which the Coulomb (Newton) properties 
are absent. Such a classification of gauge fields leads us to 
\begin{equation}
\gamma_{E}^{C}, \, \, \, \, \gamma_{W}^{C}, \, \, \, \, \gamma_{S}^{C}, \, \, \, \, ...\rightarrow 
\vec{\gamma}_{E}^{C}, \, \, \, \, \vec{\gamma}_{W}^{C}, \, \, \, \, 
\vec{\gamma}_{S}^{C}, \, \, \, \, ...,
\label{79}
\end{equation}
\begin{equation}
\gamma_{E}^{N}, \, \, \, \, \gamma_{W}^{N}, \, \, \, \, \gamma_{S}^{N}, \, \, \, \, ...\rightarrow 
\vec{\gamma}_{E}^{N}, \, \, \, \, \vec{\gamma}_{W}^{N}, \, \, \, \, 
\vec{\gamma}_{S}^{N}, \, \, \, \, ...,
\label{80}
\end{equation}
\begin{equation}
\gamma_{H_{E}}^{H_{C}}, \, \, \, \, \gamma_{H_{W}}^{H_{C}}, \, \, \, \,
\gamma_{H_{S}}^{H_{C}}, \, \, \, \, ...\rightarrow
\vec{\gamma}_{H_{E}}^{H_{C}}, \, \, \, \, \vec{\gamma}_{H_{W}}^{H_{C}}, \, \, \, \,
\vec{\gamma}_{H_{S}}^{H_{C}}, \, \, \, \, ...,
\label{81}
\end{equation}
\begin{equation}
\gamma_{H_{E}}^{H_{N}}, \, \, \, \, \gamma_{H_{W}}^{H_{N}}, \, \, \, \,
\gamma_{H_{S}}^{H_{N}}, \, \, \, \, ...\rightarrow
\vec{\gamma}_{H_{E}}^{H_{N}}, \, \, \, \, \vec{\gamma}_{H_{W}}^{H_{N}}, \, \, \, \,
\vec{\gamma}_{H_{S}}^{H_{N}}, \, \, \, \, ....
\label{82}
\end{equation}

Presentations (\ref{79}) and (\ref{81}) feature the idea that each type of Coulomb $\vec{\gamma}_{K}^{C}$ field testifies in favor of the existence of a kind of magnetocoulomb $\vec{\gamma}_{H_{K}}^{H_{C}}$ field. Such pairs constitute the sequence of allcoulomb fields (\ref{75}), the first term of which was always accepted as an electromagnetic field, and the roles 
of other terms in the establishment of a true picture of grand unification have remained hitherto latent. We must therefore express the mathematical connections between $K$ and $H_{K}$ elements 
of an allcoulomb triple of subsets
\begin{equation}
\vec{\gamma}_{KH_{K}}^{C}=\{\vec{\gamma}_{K}^{C}, \, \, \, \, 
\vec{\gamma}_{H_{K}}^{H_{C}}\}, \, \, \, \,
\rho_{KH_{K}}^{C}=\{\rho_{K}^{C}, \, \, \, \, \rho_{H_{K}}^{H_{C}}\}, \, \, \, \, 
J_{KH_{K}}^{C}=\{J_{K}^{C}, \, \, \, \, J_{H_{K}}^{H_{C}}\}
\label{83}
\end{equation}
of an allcoulomb field $\vec{\gamma}_{KH_{K}}^{C}$ and a density $\rho_{KH_{K}}^{C}$ of charge
$e_{\gamma_{KH_{K}}}^{K}$ of an allcoulomb boson $\gamma_{KH_{K}}^{C}$ and its current 
$J_{KH_{K}}^{C}$ in the following manner:
\begin{equation}
div\vec{\gamma}_{K}^{C}=4\pi\rho_{K}^{C},
\label{84}
\end{equation}
\begin{equation}
rot\vec{\gamma}_{H_{K}}^{H_{C}}=\frac{4\pi}{c}J_{K}^{C}+
\frac{1}{c}\frac{\partial\vec{\gamma}_{K}^{C}}{\partial t},
\label{85}
\end{equation} 
\begin{equation}
div\vec{\gamma}_{H_{K}}^{H_{C}}=4\pi\rho_{H_{K}}^{H_{C}},
\label{86}
\end{equation} 
\begin{equation}
rot\vec{\gamma}_{K}^{C}=\frac{4\pi}{c}J_{H_{K}}^{H_{C}}-
\frac{1}{c}\frac{\partial\vec{\gamma}_{H_{K}}^{H_{C}}}{\partial t}.
\label{87}
\end{equation} 

In the first $K$ system of equations, there is a density $\rho_{K}^{C}$ of Coulomb charge
$e_{\gamma_{K}}^{K}$ of boson $\gamma_{K}^{C}$ with a Coulomb $J_{K}^{C}$ current. In the 
first $H_{K}$ system of equations appears a density $\rho_{H_{K}}^{H_{C}}$ of magnetocoulomb 
charge $e_{\gamma_{H_{K}}}^{H_{K}}$ of monoboson $\gamma_{H_{K}}^{H_{C}}$ with a magnetocoulomb
$J_{H_{K}}^{H_{C}}$ current. 

On this basis, at $K=E$ and $H_{K}=H_{E},$ Dirac [42,43] connected a Coulomb charge of one boson 
with a magnetocoulomb charge of its own monoboson. However, the question of their existence 
as a photon-monophoton pair [15] was not raised. 

Presentations (\ref{80}) and (\ref{82}) express the regularity that each Newton $\vec{\gamma}_{K}^{N}$ field has a magnetonewton $ \vec{\gamma}_{H_{K}}^{H_{N}}$ field. These pairs constitute the sequence of allnewton fields (\ref{76}), none of which was known before the creation of the first electromagnetic theory. Thereby, one must establish the mathematical connections between $K$ and $H_{K}$ elements of an allnewton triple of subsets
\begin{equation}
\vec{\gamma}_{KH_{K}}^{N}=\{\vec{\gamma}_{K}^{N}, \, \, \, \, 
\vec{\gamma}_{H_{K}}^{H_{N}}\}, \, \, \, \,
\rho_{KH_{K}}^{N}=\{\rho_{K}^{N}, \, \, \, \, \rho_{H_{K}}^{H_{N}}\}, \, \, \, \, 
J_{KH_{K}}^{N}=\{J_{K}^{N}, \, \, \, \, J_{H_{K}}^{H_{N}}\}
\label{88}
\end{equation}
of an allnewton field $\vec{\gamma}_{KH_{K}}^{N}$ and a density $\rho_{KH_{K}}^{N}$ of mass
$m_{\gamma_{KH_{K}}}^{K}$ of an allnewton boson $\gamma_{KH_{K}}^{N}$ and its current $J_{KH_{K}}^{N}$ in an explicit form as         
\begin{equation}
div\vec{\gamma}_{K}^{N}=4\pi\rho_{K}^{N},
\label{89}
\end{equation}
\begin{equation}
rot\vec{\gamma}_{H_{K}}^{H_{N}}=\frac{4\pi}{c}J_{K}^{N}+
\frac{1}{c}\frac{\partial\vec{\gamma}_{K}^{N}}{\partial t},
\label{90}
\end{equation} 
\begin{equation}
div\vec{\gamma}_{H_{K}}^{H_{N}}=4\pi\rho_{H_{K}}^{H_{N}},
\label{91}
\end{equation} 
\begin{equation}
rot\vec{\gamma}_{K}^{N}=\frac{4\pi}{c}J_{H_{K}}^{H_{N}}-
\frac{1}{c}\frac{\partial\vec{\gamma}_{H_{K}}^{H_{N}}}{\partial t}.
\label{92}
\end{equation} 

The second $K$ system of equations contains a density $\rho_{K}^{N}$ of Newton mass 
$m_{\gamma_{K}}^{K}$ of boson $\gamma_{K}^{N}$ with a Newton $J_{K}^{N}$ current. The second 
$H_{K}$ system of equations includes a density $\rho_{H_{K}}^{H_{N}}$ of magnetonewton mass  $m_{\gamma_{H_{K}}}^{H_{K}}$ of monoboson $\gamma_{H_{K}}^{H_{N}}$ with a magnetonewton $J_{H_{K}}^{H_{N}}$ current. 

So it is seen that nature itself is not in force to characterize the same Coulomb (Newton) boson regardless of its own magnetocoulomb (magnetonewton) monoboson. It hereby relates a Newton mass 
of each boson even in the case of an allnewton system to the corresponding magnetonewton mass 
of its own monoboson.

Furthermore, if graviton $\gamma_{GU}^{C} (\gamma_{GU}^{N})$ unites bosons 
$\gamma_{K}^{C} (\gamma_{K}^{N})$ in a unified whole as the mediating boson of the gravitational 
$(GU)$ interaction Coulomb (Newton) component, the gravitons (\ref{50}) and (\ref{51}) replace (\ref{79}) and (\ref{80}) for
\begin{equation}
\gamma_{GU}^{C}=\{\gamma_{E}^{C}, \, \, \, \, \gamma_{W}^{C}, \, \, \, \, 
\gamma_{S}^{C}, \, \, \, \, ...\}\rightarrow \vec{\gamma}_{GU}^{C}=
\{\vec{\gamma}_{E}^{C}, \, \, \, \, \vec{\gamma}_{W}^{C}, \, \, \, \, 
\vec{\gamma}_{S}^{C}, \, \, \, \, ...\},
\label{93}
\end{equation}
\begin{equation}
\gamma_{GU}^{N}=\{\gamma_{E}^{N}, \, \, \, \, \gamma_{W}^{N}, \, \, \, \, 
\gamma_{S}^{N}, \, \, \, \, ...\}\rightarrow \vec{\gamma}_{GU}^{N}=
\{\vec{\gamma}_{E}^{N}, \, \, \, \, \vec{\gamma}_{W}^{N}, \, \, \, \, 
\vec{\gamma}_{S}^{N}, \, \, \, \, ...\}.
\label{94}
\end{equation}

Without contradicting the symmetry laws, monobosons $\gamma_{H_{K}}^{H_{C}} (\gamma_{H_{K}}^{H_{N}})$ constitute a single monograviton $\gamma_{GU_{H}}^{H_{C}} (\gamma_{GU_{H}}^{H_{N}})$ as the mediating monoboson of the magnetogravitational $(GU_{H})$ interaction magnetocoulomb (magnetonewton) component. The monogravitons (\ref{57}) and (\ref{58}) thus formed transform (\ref{81}) and (\ref{82}) into 
\begin{equation}
\gamma_{GU_{H}}^{H_{C}}=
\{\gamma_{H_{E}}^{H_{C}}, \, \, \, \, \gamma_{H_{W}}^{H_{C}}, \, \, \, \,
\gamma_{H_{S}}^{H_{C}}, \, \, \, \, ...\}\rightarrow \vec{\gamma}_{GU_{H}}^{H_{C}}=
\{\vec{\gamma}_{H_{E}}^{H_{C}}, \, \, \, \, \vec{\gamma}_{H_{W}}^{H_{C}}, \, \, \, \,
\vec{\gamma}_{H_{S}}^{H_{C}}, \, \, \, \, ...\},
\label{95}
\end{equation}
\begin{equation}
\gamma_{GU_{H}}^{H_{N}}=
\{\gamma_{H_{E}}^{H_{N}}, \, \, \, \, \gamma_{H_{W}}^{H_{N}}, \, \, \, \,
\gamma_{H_{S}}^{H_{N}}, \, \, \, \, ...\}\rightarrow \vec{\gamma}_{GU_{H}}^{H_{N}}=
\{\vec{\gamma}_{H_{E}}^{H_{N}}, \, \, \, \, \vec{\gamma}_{H_{W}}^{H_{N}}, \, \, \, \,
\vec{\gamma}_{H_{S}}^{H_{N}}, \, \, \, \, ...\}.
\label{96}
\end{equation}

These connections clearly show that an allcoulomb field 
\begin{equation}
\{\vec{\gamma}_{GU}^{C}, \, \, \, \, \vec{\gamma}_{GU_{H}}^{H_{C}}\}=
\{\{\vec{\gamma}_{E}^{C}, \, \, \, \, \vec{\gamma}_{H_{E}}^{H_{C}}\}, \, \, \, \,
\{\vec{\gamma}_{W}^{C}, \, \, \, \, \vec{\gamma}_{H_{W}}^{H_{C}}\}, \, \, \, \,
\{\vec{\gamma}_{S}^{C}, \, \, \, \, \vec{\gamma}_{H_{S}}^{H_{C}}\}, \, \, \, \, ...\}
\label{97}
\end{equation}
corresponds to a kind of allnewton field
\begin{equation}
\{\vec{\gamma}_{GU}^{N}, \, \, \, \, \vec{\gamma}_{GU_{H}}^{H_{N}}\}=
\{\{\vec{\gamma}_{E}^{N}, \, \, \, \, \vec{\gamma}_{H_{E}}^{H_{N}}\}, \, \, \, \,
\{\vec{\gamma}_{W}^{N}, \, \, \, \, \vec{\gamma}_{H_{W}}^{H_{N}}\}, \, \, \, \,
\{\vec{\gamma}_{S}^{N}, \, \, \, \, \vec{\gamma}_{H_{S}}^{H_{N}}\}, \, \, \, \, ...\}.
\label{98}
\end{equation}

This one-to-one correspondence between the fields of allcoulomb and allnewton gravitons is crucial for establishing all mathematical connections within each pair of allgravitational fields.

To show this, one must refer to the structure and component of an allcoulomb triple 
of subsets
\begin{equation}
\vec{\gamma}_{AG}^{C}=\{\vec{\gamma}_{GU}^{C}, \, \, \, \, 
\vec{\gamma}_{GU_{H}}^{H_{C}}\}, \, \, \, \,
\rho_{AG}^{C}=\{\rho_{GU}^{C}, \, \, \, \, \rho_{GU_{H}}^{H_{C}}\}, \, \, \, \, 
J_{AG}^{C}=\{J_{GU}^{C}, \, \, \, \, J_{GU_{H}}^{H_{C}}\}
\label{99}
\end{equation}
of an allcoulomb field $\vec{\gamma}_{AG}^{C}$ and a density $\rho_{AG}^{C}$ of charge $e_{\gamma_{AG}}^{K}$ of an allcoulomb graviton $\gamma_{AG}^{C}$ and its current $J_{AG}^{C}$ 
because the mathematical connections between their $K$ and $H_{K}$ elements unite the curved 
vector fields $\vec{\gamma}_{GU}^{C}$ and $\vec{\gamma}_{GU_{H}}^{H_{C}}$ in a unified whole 
as the field of the mediating boson 
\begin{equation}
\gamma_{AG}^{C}=\{\gamma_{GU}^{C}, \, \, \, \, \gamma_{GU_{H}}^{H_{C}}\}
\label{100}
\end{equation}
of an allgravitational $(AG)$ interaction allcoulomb component
\begin{equation}
div\vec{\gamma}_{GU}^{C}=4\pi\rho_{GU}^{C},
\label{101}
\end{equation}
\begin{equation}
rot\vec{\gamma}_{GU_{H}}^{H_{C}}=\frac{4\pi}{c}J_{GU}^{C}+
\frac{1}{c}\frac{\partial\vec{\gamma}_{GU}^{C}}{\partial t},
\label{102}
\end{equation} 
\begin{equation}
div\vec{\gamma}_{GU_{H}}^{H_{C}}=4\pi\rho_{GU_{H}}^{H_{C}},
\label{103}
\end{equation} 
\begin{equation}
rot\vec{\gamma}_{GU}^{C}=\frac{4\pi}{c}J_{GU_{H}}^{H_{C}}-
\frac{1}{c}\frac{\partial\vec{\gamma}_{GU_{H}}^{H_{C}}}{\partial t}.
\label{104}
\end{equation} 

In the third $K$ system, $\rho_{GU}^{C}$ denotes the density of Coulomb charge $e_{\gamma_{GU}}^{C}$
of graviton $\gamma_{GU}^{C}$ and its Coulomb $J_{GU}^{C}$ current. In the third $H_{K}$ system, $\rho_{GU_{H}}^{H_{C}}$ describes the density of magnetocoulomb charge $e_{\gamma_{GU_{H}}}^{H_{C}}$ of monograviton $\gamma_{GU_{H}}^{H_{C}}$ and its magnetocoulomb $J_{GU_{H}}^{H_{C}}$ current. 

Of course, an observed regularity reflects the characteristic features of the structure of the Coulomb graviton and magnetocoulomb monograviton and thereby opens in principle the possibility 
for establishing the mathematical connection between their charges.

In a similar way, one can form an allnewton triple of subsets
\begin{equation}
\vec{\gamma}_{AG}^{N}=\{\vec{\gamma}_{GU}^{N}, \, \, \, \, 
\vec{\gamma}_{GU_{H}}^{H_{N}}\}, \, \, \, \,
\rho_{AG}^{N}=\{\rho_{GU}^{N}, \, \, \, \, \rho_{GU_{H}}^{H_{N}}\}, \, \, \, \, 
J_{AG}^{N}=\{J_{GU}^{N}, \, \, \, \, J_{GU_{H}}^{H_{N}}\}
\label{105}
\end{equation}
of an allnewton field $\vec{\gamma}_{AG}^{N}$ and a density $\rho_{AG}^{N}$ of mass  $m_{\gamma_{AG}}^{K}$ of an allnewton graviton $\gamma_{AG}^{N}$ and its current $J_{AG}^{N}$ and unite the curved vector fields $\vec{\gamma}_{GU}^{N}$ and $\vec{\gamma}_{GU_{H}}^{H_{N}}$ into
a single field of the mediating boson 
\begin{equation}
\gamma_{AG}^{N}=\{\gamma_{GU}^{N}, \, \, \, \, \gamma_{GU_{H}}^{H_{N}}\}
\label{106}
\end{equation}
of an allgravitational $(AG)$ interaction allnewton component
\begin{equation}
div\vec{\gamma}_{GU}^{N}=4\pi\rho_{GU}^{N},
\label{107}
\end{equation}
\begin{equation}
rot\vec{\gamma}_{GU_{H}}^{H_{N}}=\frac{4\pi}{c}J_{GU}^{N}+
\frac{1}{c}\frac{\partial\vec{\gamma}_{GU}^{N}}{\partial t},
\label{108}
\end{equation} 
\begin{equation}
div\vec{\gamma}_{GU_{H}}^{H_{N}}=4\pi\rho_{GU_{H}}^{H_{N}},
\label{109}
\end{equation} 
\begin{equation}
rot\vec{\gamma}_{GU}^{N}=\frac{4\pi}{c}J_{GU_{H}}^{H_{N}}-
\frac{1}{c}\frac{\partial\vec{\gamma}_{GU_{H}}^{H_{N}}}{\partial t}.
\label{110}
\end{equation} 

The fourth $K$ system includes a density $\rho_{GU}^{N}$ of Newton mass $m_{\gamma_{GU}}^{N}$ of graviton $\gamma_{GU}^{N}$ with a Newton $J_{GU}^{N}$ current. The fourth $H_{K}$ system contains a density $\rho_{GU_{H}}^{H_{N}}$ of magnetonewton mass  $m_{\gamma_{GU_{H}}}^{H_{N}}$ of monograviton $\gamma_{GU_{H}}^{H_{N}}$ with a magnetonewton $J_{GU_{H}}^{H_{N}}$ current. 

Of course, such a regularity reflects the sharply expressed features of the structure of the 
Newton graviton and magnetonewton monograviton and thereby opens the chance for establishing 
the mathematical connection between their masses.

It is not excluded, however, that regardless of the maximal quantity of forces and existing types of actions, the Coulomb (magnetocoulomb) and Newton (magnetonewton) gravitons (monogravitons) constitute the Newton-Coulomb (magnetonewton-magnetocoulomb) pair of gauge bosons such that it comes forward in nature as one single graviton (monograviton). Therefore, from (\ref{52}) and (\ref{56}), we are led to the following relationships: 
\begin{equation}
\gamma_{GU}=\{\gamma_{GU}^{C}, \, \, \, \, \gamma_{GU}^{N}\}\rightarrow
\vec{\gamma}_{GU}=\{\vec{\gamma}_{GU}^{C}, \, \, \, \, \vec{\gamma}_{GU}^{N}\},
\label{111}
\end{equation}
\begin{equation}
\gamma_{GU_{H}}=\{\gamma_{GU_{H}}^{H_{C}}, \, \, \, \, \gamma_{GU_{H}}^{H_{N}}\}\rightarrow
\vec{\gamma}_{GU_{H}}=\{\vec{\gamma}_{GU_{H}}^{H_{C}}, \, \, \, \, \vec{\gamma}_{GU_{H}}^{H_{N}}\}.
\label{112}
\end{equation}

They together with fields (\ref{97}) and (\ref{98}) show that there exists a range of structural connections between graviton $\gamma_{GU}$ and monograviton $\gamma_{GU_{H}},$ which unite them in an allgraviton $\gamma_{AG},$ namely, in a mediating boson (\ref{66}) of a grand united allgravitational $(AG)$ interaction.

Furthermore, if it turns out that 
\begin{equation}
\gamma_{AG}=\{\gamma_{GU}, \, \, \, \, \gamma_{GU_{H}}\}\rightarrow
\vec{\gamma}_{AG}=\{\vec{\gamma}_{GU}, \, \, \, \, \vec{\gamma}_{GU_{H}}\},
\label{113}
\end{equation}
from the point of view of each pair of fields (\ref{111}) or (\ref{112}), the structural fields $\vec{\gamma}_{AG}$ of an allgraviton $\gamma_{AG}$ should be chosen in the form
\begin{equation}
\{\vec{\gamma}_{GU}, \, \, \, \, \vec{\gamma}_{GU_{H}}\}\rightarrow
\{\{\vec{\gamma}_{GU}^{C}, \, \, \, \, \vec{\gamma}_{GU}^{N}\}, \, \, \, \,
\{\vec{\gamma}_{GU_{H}}^{H_{C}}, \, \, \, \, \vec{\gamma}_{GU_{H}}^{H_{N}}\}\}.
\label{114}
\end{equation}

The mathematical connections come forward in it as 
\begin{equation}
div\vec{\gamma}_{GU}=4\pi\rho_{GU},
\label{115}
\end{equation}
\begin{equation}
rot\vec{\gamma}_{GU_{H}}=\frac{4\pi}{c}J_{GU}+
\frac{1}{c}\frac{\partial\vec{\gamma}_{GU}}{\partial t},
\label{116}
\end{equation} 
\begin{equation}
div\vec{\gamma}_{GU_{H}}=4\pi\rho_{GU_{H}},
\label{117}
\end{equation} 
\begin{equation}
rot\vec{\gamma}_{GU}=\frac{4\pi}{c}J_{GU_{H}}-
\frac{1}{c}\frac{\partial\vec{\gamma}_{GU_{H}}}{\partial t}.
\label{118}
\end{equation} 

In the fifth $K$ system of equations, $\rho_{GU}$ appears as a density of charge $e_{\gamma_{GU}}$ 
of graviton $\gamma_{GU}$ and its $J_{GU}$ current. In the fifth $H_{K}$ system of equations, 
$\rho_{GU_{H}}$ arises as a density of charge $e_{\gamma_{GU_{H}}}$ of monograviton 
$\gamma_{GU_{H}}$ and its $J_{GU_{H}}$ current. 

It is clear, however, that any graviton or monograviton may serve as the source of a kind of 
charge. According to this principle, each of the equations (\ref{115})-(\ref{118}) unites one equation from (\ref{101})-(\ref{104}) with one equation from (\ref{107})-(\ref{110}) and that, consequently, each density $\rho_{AG}$ of charge $e_{\gamma_{AG}}$ of an allgraviton 
$\gamma_{AG}$ and its current $J_{AG}$ must have its own set corresponding in it to a kind of set $\vec{\gamma}_{AG}$ of its structural fields. This expresses a great responsibility of an allgravity for the structure of each pair from $K$ and $H_{K}$ systems of wave equations of a grand unification, as well as for its behavior, including at $i,$ $j=K$ $(i\neq j)$ a unified mathematical description 
of all connections between $i$ and $H_{j}$ elements of each triple of subsets
\begin{equation}
\vec{\gamma}_{iH_{j}}^{C}=\{\vec{\gamma}_{i}^{C}, \, \, \, \, 
\vec{\gamma}_{H_{j}}^{H_{C}}\}, \, \, \, \,
\rho_{iH_{j}}^{C}=\{\rho_{i}^{C}, \, \, \, \, \rho_{H_{j}}^{H_{C}}\}, \, \, \, \, 
J_{iH_{j}}^{C}=\{J_{i}^{C}, \, \, \, \, J_{H_{j}}^{H_{C}}\},
\label{119}
\end{equation}
\begin{equation}
\vec{\gamma}_{iH_{j}}^{N}=\{\vec{\gamma}_{i}^{N}, \, \, \, \, 
\vec{\gamma}_{H_{j}}^{H_{N}}\}, \, \, \, \,
\rho_{iH_{j}}^{N}=\{\rho_{i}^{N}, \, \, \, \, \rho_{H_{j}}^{H_{N}}\}, \, \, \, \, 
J_{iH_{j}}^{N}=\{J_{i}^{N}, \, \, \, \, J_{H_{j}}^{H_{N}}\}
\label{120}
\end{equation}
within the same triple of sets
\begin{equation}
\vec{\gamma}_{AG}=\{\vec{\gamma}_{GU}, \, \, \, \, \vec{\gamma}_{GU_{H}}\}, \, \, \, \,
\rho_{AG}=\{\rho_{GU}, \, \, \, \, \rho_{GU_{H}}\}, \, \, \, \, 
J_{AG}=\{J_{GU}, \, \, \, \, J_{GU_{H}}\}.
\label{121}
\end{equation}

Thus, we need to use each of the gauge fields at the level of a grand unification mathematical logic as one of the existing types of subsets of the same set such that it comes forward in nature as the field of an allgraviton consisting of fields of gravitons and monogravitons.

Finally, insofar as the reliable practical information about graviton, monograviton, and 
allgraviton is concerned, we recognize that there is no single regularity of gravity, magnetogravity, and allgravity for revealing which of the infinitely many selected devices of natural origin would not exist. One such an object, according to our observation, is the northern magnetogravitational lights. However, here we have already mentioned that nature itself in a curved space-time of allgraviton fields. 

\vspace{0.6cm}
\noindent
{\bf References}
\begin{enumerate}
\item 
R.S. Sharafiddinov, Spacetime Subst. {\bf 3}, 47 (2002); Bull. Am. Phys. 

Soc. {\bf 59}, T1.00009 (2014); e-print arXiv:physics/0305008.
\item
R.S. Sharafiddinov, Spacetime Subst. {\bf 2}, 87 (2003); Bull. Am. Phys. 

Soc. {\bf 59}, EC.00006 (2014); e-print arXiv:hep-ph/0401230.
\item
C.L. Cowan and F. Reines, Phys. Rev. {\bf 107}, 528 (1957).

DOI: https://doi.org/10.1103/PhysRev.107.528
\item
T. Fliessbach, Phys. Rev. {\bf D 18}, 3028 (1978).

DOI: https://doi.org/10.1103/PhysRevD.18.3028
\item
M.A. Chernikov et al., Phys. Rev. Lett. {\bf 68}, 3383 (1992).

DOI: https://doi.org/10.1103/PhysRevLett.68.3383
\item
E. Fischbach et al., Phys. Rev. Lett. {\bf 73}, 514 (1994).

DOI: https://doi.org/10.1103/PhysRevLett.73.514
\item
B. Mashhoon, J.H. Paik and C.M. Will, Phys. Rev. {\bf D 39}, 2825 (1989).

DOI: https://doi.org/10.1103/PhysRevD.39.2825
\item
B. Mashhoon, Phys. Lett. {\bf A 173}, 347 (1993). 

DOI: https://doi.org/10.1016/0375-9601(93)90248-X 
\item
N. Li and D.G. Torr, Phys. Rev. {\bf D 43}, 457 (1991).

DOI:https://doi.org/10.1103/PhysRevD.43.457
\item
C.V. Weinheimer et al., Phys. Lett. {\bf B 460}, 219 (1999).

DOI: https://doi.org/10.1016/S0370-2693(99)00780-7
\item
V.M. Lobashev et al., Phys. Lett. {\bf B 460}, 227 (1999).

DOI: https://doi.org/10.1016/S0370-2693(99)00781-9
\item
E.R. Williams, J.E. Faller and H.A. Hill, Phys. Rev. Lett. {\bf 26}, 721 (1971).

DOI: https://doi.org/10.1103/PhysRevLett.26.721
\item
R.S. Sharafiddinov, Spacetime Subst. {\bf 1}, 176 (2000); e-print arXiv:hep-ph/0305009.
\item
H. Harrison et al., Am. J. Phys. {\bf 31}, 249 (1963).

DOI: https://doi.org/10.1119/1.1969426                                               
\item
R.S. Sharafiddinov, Spacetime Subst. {\bf 3}, 132 (2002); Bull. Am. Phys. 

Soc. {\bf 59}, T1.00005 (2014); e-print arXiv:physics/0305014.
\item
G. Raffelt, Phys. Rev. {\bf D 50}, 7729 (1994).

DOI: https://doi.org/10.1103/PhysRevD.50.7729
\item
R.S. Sharafiddinov, Can. J. Phys. {\bf 92}, 1262 (2014). 

DOI: https://doi.org/10.1139/cjp-2013-0458
\item
I.Z. Fisher, Zh. Eksp. Teor. Fiz. {\bf 20}, 956 (1950).
\item
R.S. Sharafiddinov, Spacetime Subst. {\bf 3}, 86 (2002); Bull. Am. Phys. 

Soc. {\bf 60}, T1.00033 (2015); e-print arXiv:physics/0305009.
\item 
R.S. Sharafiddinov, Spacetime Subst. {\bf 5}, 83 (2004); e-print arXiv:hep-ph/0306255.
\item
R.S. Sharafiddinov, Phys. Essays {\bf 29}, 410 (2016). 

DOI: https://doi.org/10.4006/0836-1398-29.3.410
\item
R.S. Sharafiddinov, Bull. Am. Phys. Soc. {\bf 60}, E13.00008 (2015); 

e-print arXiv:hep-ph/0409254v4.
\item
R.S. Sharafiddinov, Phys. Essays {\bf 32}, 358 (2019). 

DOI: https://dx.doi.org/10.4006/0836-1398-32.3.358
\item
H. Bondi, Rev. Mod. Phys. {\bf 29}, 423 (1957).

DOI: https://doi.org/10.1103/RevModPhys.29.423
\item
B.D. Miller, Astrophys. J. {\bf 208}, 275 (1976). 
\item
H.A. Aspden, Phys. Essays {\bf 4}, 13 (1991).
\item
R.S. Sharafiddinov, Spacetime Subst. {\bf 3}, 134 (2002); Bull. Am. Phys. 

Soc. {\bf 61}, T1.00020 (2016); e-print arXiv:physics/0305015.
\item
S.L. Glashow, Nucl. Phys. {\bf 22}, 579 (1961).

DOI: https://doi.org/10.1016/0029-5582(61)90469-2 
\item
A. Salam and J.C. Ward, Phys. Lett. {\bf 13}, 168 (1964).

DOI: https://doi.org/10.1016/0031-9163(64)90711-5
\item
S. Weinberg, Phys. Rev. Lett. {\bf 19}, 1264 (1967).

DOI: https://doi.org/10.1103/PhysRevLett.19.1264
\item
R.S. Sharafiddinov, Can. J. Phys. {\bf 93}, 1005 (2015). 

DOI: https://doi.org/10.1139/cjp-2014-0497
\item
R.S. Sharafiddinov, Phys. Essays {\bf 19}, 58 (2006).

DOI: https://doi.org/10.4006/1.3025781
\item
R.S. Sharafiddinov, Bull. Am. Phys. Soc. {\bf 59}, T1.00004 (2014);

e-print arXiv:hep-ph/0511065.
\item
J.C. Pati and A. Salam, Phys. Rev. {\bf D 10}, 275 (1974).

DOI: https://doi.org/10.1103/PhysRevD.10.275
\item
L.B. Okun, Yad. Phys. {\bf 45}, 158 (1987).
\item
P.W. Higgs, Phys. Rev. Lett. {\bf 13}, 508 (1964).

DOI: https://doi.org/10.1103/PhysRevLett.13.508
\item
A.A. Michelson, Am. J. Sci. {\bf 22}, 120 (1881).

DOI: https://doi.org/10.2475/ajs.s3-22.128.120 
\item
A.A. Michelson and E.W. Morley, Am. J. Sci. {\bf 34}, 333 (1887).

DOI: https://doi.org/10.2475/ajs.s3-34.203.333 
\item
A.A. Michelson and E.W. Morley, Philos. Mag. {\bf 24}, 449 (1887).

DOI: https://doi.org/10.1080/14786448708628130
\item
T.D. Lee and C.N. Yang, Phys. Rev. {\bf 105}, 1671 (1957).

DOI: https://doi.org/10.1103/PhysRev.105.1671
\item
R.S. Sharafiddinov, Bull. Am. Phys. Soc. {\bf 60}, D1.00036 (2015); 

e-print arXiv:hep-ph/0409254v4.
\item
P.A.M. Dirac, Proc. Roy. Soc. {\bf A 133}, 60 (1931).

DOI: https://doi.org/10.1098/rspa.1931.0130
\item
P.A.M. Dirac, Phys. Rev. {\bf D 74}, 817 (1948).

DOI: https://doi.org/10.1103/PhysRev.74.817
\end{enumerate}
\end{document}